\documentclass{aastex}
\usepackage{amsmath}
\usepackage{graphicx}
\usepackage{url}
\usepackage{amssymb}
\usepackage{float}
\makeatletter
\let\@dates\relax
\makeatother
\shortauthors{ADRIANI ET AL.}

\shorttitle{PAMELA'S GEOMAGNETIC CUTOFF MEASUREMENTS}

\begin{document}

\title{PAMELA's measurements of geomagnetic cutoff variations during the 14 December 2006 storm}

\author{
O.~Adriani$^{1,2}$,
G.~C.~Barbarino$^{3,4}$,
G.~A.~Bazilevskaya$^{5}$,
R.~Bellotti$^{6,7}$,
M.~Boezio$^{8}$,
E.~A.~Bogomolov$^{9}$,
M.~Bongi$^{1,2}$,
V.~Bonvicini$^{8}$,
S.~Bottai$^{2}$,
A.~Bruno$^{6,7,*}$,
F.~Cafagna$^{7}$,
D.~Campana$^{4}$,
P.~Carlson$^{10}$,
M.~Casolino$^{11,12}$,
G.~Castellini$^{13}$,
C.~De~Donato$^{11,14}$,
G.~A.~de~Nolfo$^{15}$,
C.~De~Santis$^{11,14}$,
N.~De~Simone$^{11}$,
V.~Di~Felice$^{11,16}$,
A.~M.~Galper$^{17}$,
A.~V.~Karelin$^{17}$,
S.~V.~Koldashov$^{17}$,
S.~Koldobskiy$^{17}$,
S.~Y.~Krutkov$^{9}$,
A.~N.~Kvashnin$^{5}$,
A.~Leonov$^{17}$,
V.~Malakhov$^{17}$,
L.~Marcelli$^{11,14}$,
M.~Martucci$^{14,18}$,
A.~G.~Mayorov$^{17}$,
W.~Menn$^{19}$,
M.~Merg$\acute{e}$$^{11,14}$,
V.~V.~Mikhailov$^{17}$,
E.~Mocchiutti$^{8}$,
A.~Monaco$^{6,7}$,
N.~Mori$^{1,2}$,
R.~Munini$^{8,20}$,
G.~Osteria$^{4}$,
F.~Palma$^{11,14}$,
B.~Panico$^{4}$,
P.~Papini$^{2}$,
M.~Pearce$^{10}$,
P.~Picozza$^{14}$,
M.~Ricci$^{18}$,
S.~B.~Ricciarini$^{2,13}$,
R.~Sarkar$^{21,22}$,
V.~Scotti$^{3,4}$,
M.~Simon$^{19}$,
R.~Sparvoli$^{11,14}$,
P.~Spillantini$^{17,23}$,
Y.~I.~Stozhkov$^{5}$,
A.~Vacchi$^{8}$,
E.~Vannuccini$^{2}$,
G.~I.~Vasilyev$^{9}$,
S.~A.~Voronov$^{17}$,
Y.~T.~Yurkin$^{17}$,
G.~Zampa$^{8}$
and N.~Zampa$^{8}$
}

\affil{$^{1}$ Department of Physics and Astronomy, University of Florence, I-50019 Sesto Fiorentino, Florence, Italy.}
\affil{$^{2}$ INFN, Sezione di Florence, I-50019 Sesto Fiorentino, Florence, Italy.}
\affil{$^{3}$ Department of Physics, University of Naples ``Federico II'', I-80126 Naples, Italy.}
\affil{$^{4}$ INFN, Sezione di Naples, I-80126 Naples, Italy.}
\affil{$^{5}$ Lebedev Physical Institute, RU-119991 Moscow, Russia}
\affil{$^{6}$ Department of Physics, University of Bari, I-70126 Bari, Italy.}
\affil{$^{7}$ INFN, Sezione di Bari, I-70126 Bari, Italy.}
\affil{$^{8}$ INFN, Sezione di Trieste, I-34149 Trieste, Italy.}
\affil{$^{9}$ Ioffe Physical Technical Institute, RU-194021 St. Petersburg, Russia.}
\affil{$^{10}$ KTH, Department of Physics, and the Oskar Klein Centre for Cosmoparticle Physics, AlbaNova University Centre, SE-10691 Stockholm, Sweden.}
\affil{$^{11}$ INFN, Sezione di Rome ``Tor Vergata'', I-00133 Rome, Italy.}
\affil{$^{12}$ RIKEN, Advanced Science Institute, Wako-shi, Saitama, Japan.}
\affil{$^{13}$ IFAC, I-50019 Sesto Fiorentino, Florence, Italy.}
\affil{$^{14}$ Department of Physics, University of Rome ``Tor Vergata'', I-00133 Rome, Italy.}
\affil{$^{15}$ Heliophysics Division, NASA Goddard Space Flight Center, Greenbelt, MD, USA.}
\affil{$^{16}$ Agenzia Spaziale Italiana (ASI) Science Data Center, Via del Politecnico snc, I-00133 Rome, Italy.}
\affil{$^{17}$ National Research Nuclear University MEPhI, RU-115409 Moscow, Russia.}
\affil{$^{18}$ INFN, Laboratori Nazionali di Frascati, Via Enrico Fermi 40, I-00044 Frascati, Italy.}
\affil{$^{19}$ Department of Physics, Universit\"{a}t Siegen, D-57068 Siegen, Germany.}
\affil{$^{20}$ Department of Physics, University of Trieste, I-34147 Trieste, Italy.}
\affil{$^{21}$ Indian Centre for Space Physics, 43 Chalantika, Garia Station Road, Kolkata 700084, West Bengal, India.}
\affil{$^{22}$ Formerly at INFN, Sezione di Trieste, I-34149 Trieste, Italy.}
\affil{$^{23}$ IAPS/INAF, I-00133 Rome, Italy.}
\affil{* Corresponding author. E-mail address: alessandro.bruno@ba.infn.it.}

\begin{abstract}
Data from the Payload for Antimatter Matter Exploration and Light-nuclei Astrophysics (PAMELA) satellite experiment were used to measure the geo\-ma\-gne\-tic cutoff for high-energy ($\gtrsim$ 80 MeV) protons during the 14 December 2006 geomagnetic storm. The variations of the cutoff latitude as a function of rigidity were studied on relatively short timescales, corresponding to spacecraft orbital periods ($\sim$94 min). Estimated cutoff values were compared with those obtained by means of a trajectory tracing approach based on a dynamical empirical mo\-de\-ling of the Earth's magnetosphere. We found significant variations in the cutoff latitude, with a maximum suppression of $\sim$7 deg at lowest rigidities during the main phase of the storm. The observed reduction in the geomagnetic shielding and its temporal evolution were related to the changes in the magnetospheric configuration, investi\-ga\-ting the role of interplanetary magnetic field, solar wind and geomagnetic parameters.
PAMELA's results represent the first direct measurement of geo\-ma\-gnetic cutoffs for protons with kinetic energies in the sub-GeV and GeV region.
\end{abstract}

%\begin{article}

\section{Introduction}
The Cosmic Ray (CR) access to a specific location in the Earth's magnetosphere is determined by the spatial structure and intensity of the geomagnetic field \citep{STORMER,SMART2001},
which is a highly dynamical system: its configuration is driven by the Solar Wind (SW) and by the interaction between the terrestrial and interplanetary fields,
being compressed at the dayside and stretched toward the magnetotail on the nightside.

Major space weather phenomena are caused by Solar Energetic Particle (SEP) events associated with explosive processes occurring in the solar atmosphere.
SEPs can si\-gni\-fi\-cantly increase the radiation dose rates compared with geomagnetically quiet times, disturbing satellite operations and producing hazardous effects to manned and robotic flight missions in the near-Earth environment, including aircrafts with their crew and passengers \citep{DYER}, and influencing the atmospheric chemistry and dynamics \citep{DANILOV}.

Large SEP events can strongly perturb the magnetosphere. In case of earthward-directed Coronal Mass Ejections (CMEs) or Co-rotating Interaction Regions (CIRs),
disturbances can culminate in geomagnetic storms, characterized by a large transfer of the SW energy into the Earth's magnetosphere, with significant changes in the currents, plasmas and fields \citep{LESKE2001}.
The reconnection of the field lines is more efficient when the Inter\-pla\-ne\-ta\-ry Ma\-gne\-tic Field (IMF) is antiparallel to the terrestrial field on the dayside boundary of the magnetosphere \citep{DUNGEY,AKASOFU,RUSSELL}. Intense geomagnetic storms can reduce the geomagnetic shielding and thus affect the planetary CR distribution \citep{DORMAN,FLUECKIGER1986,SMART2000}.

An adequate description of the geomagnetic cutoff during SEP events has been the object of several studies based on
multiple approaches, including spacecraft \citep{MAZUR1999,LESKE2001,OGLIORE,BIRCH} and ground-based \citep{RODGER2006,TYASTO2013} observations,
and calculations mainly based on tracing particles through models of the geomagnetic field \citep{SMART1985,SMART2001,SMART2003,Kress2010}.
Simplified empirical cutoff models have been developed, by parameterizing observations in terms of the $Kp$ or $Dst$ indices \citep{LESKE2001,BIRCH,NEAL2013} or using multi-variable approaches \citep{DMITRIEV2010}.

PAMELA's measurements of relatively quiet magnetospheric cutoffs can be found in publications \citep{ALBEDO}.
In this work we present the measurement of the cutoff variability during the strong geomagnetic storm on 14 December 2006,
the last large CME-driven storm of the 23$^{rd}$ solar cycle.

\begin{figure*}[t!]
\centering
\includegraphics[width=5.3in]{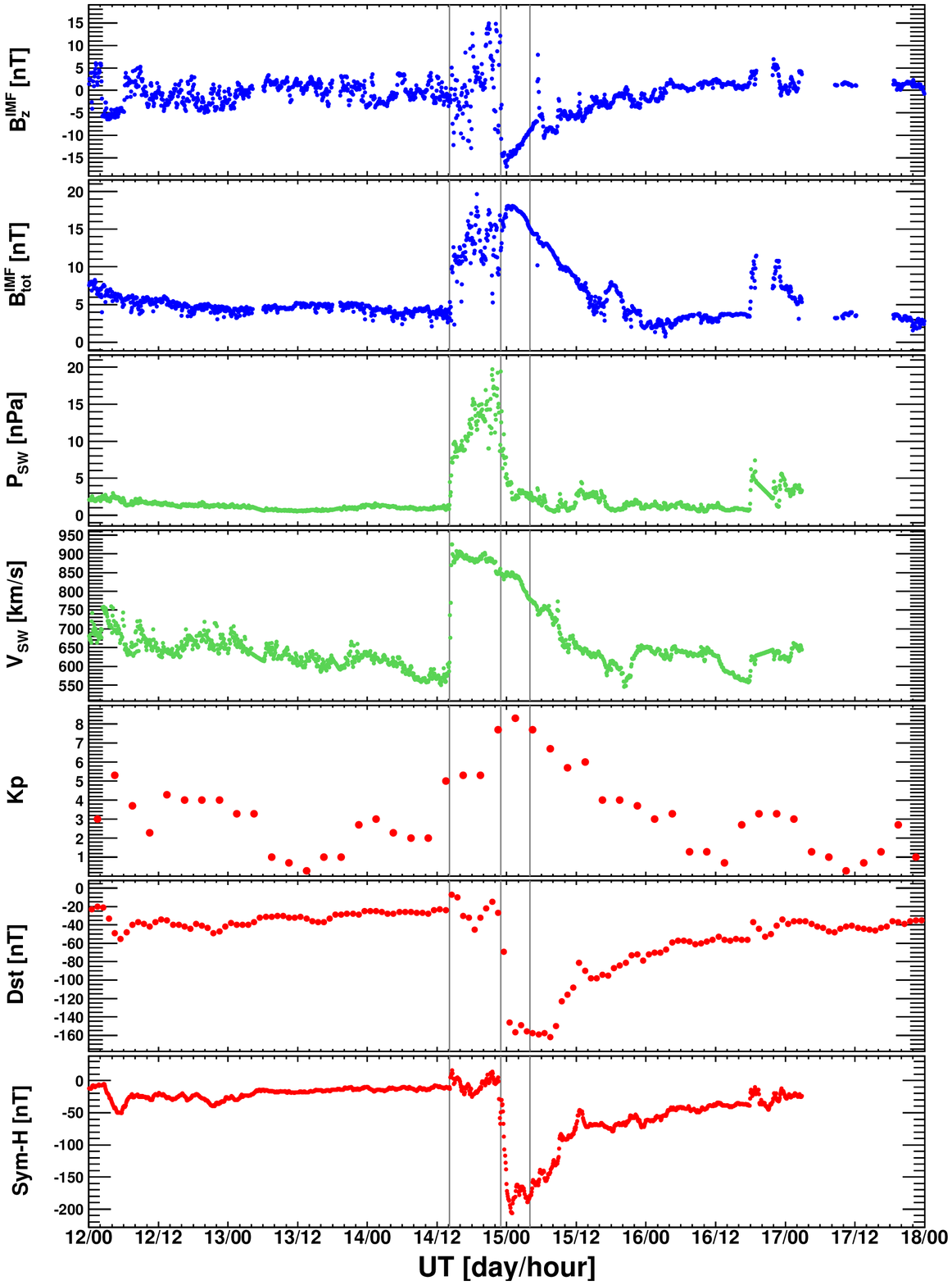}
\caption{Time profiles of the main IMF ($B_{z}^{IMF}$ and $B_{tot}^{IMF}$; blue), solar wind ($P_{SW}$ and $V_{SW}$; green) and geomagnetic ($Kp$, $Dst$, $Sym$-$H$; red) parameters between 12--18 December 2006. The vertical lines mark the beginning of the storm initial, main and recovery phases, respectively.}
\label{interplanetary_params}
\end{figure*}

\section{The 14 December 2006 Geomagnetic Storm}\label{The 14 December 2006 Geomagnetic Storm}
On 13 December 2006 at 0214 UT, an X3.4/4B solar flare occurred in the active region NOAA 10930 (S06W23).
This event also produced a full-halo CME with a sky plane projected speed of 1774 km s$^{-1}$ (\url{http://cdaw.gsfc.nasa.gov/CME_list/}).
The forward shock of the CME reached the Earth at about 1410 UT on 14 December causing a Forbush decrease of Galactic CR intensities that lasted for several days.
The X1.5 flare (S06W46) at 2107 UT on 14 December gave start to a new growth of particle intensity, as recorded by PAMELA and other space-based detectors.
The corresponding CME had a velocity of 1042 km s$^{-1}$. PAMELA's measurements of the proton and helium fluxes during the 13--14 December 2006 events can be found in publications \citep{SEP2006}.

Figure \ref{interplanetary_params} reports the variations (12--18 December 2006) in the main IMF, SW and geomagnetic parameters.
SW and IMF values were obtained from the high resolution (5-min) Omniweb database (\url{http://omniweb.gsfc.nasa.gov}), which provides in-situ observations time-shifted to the bow shock nose of the Earth \citep{OMNIWEB}. In particular, the interplanetary data for the considered time period are based on the measurements of the Advanced Composition Explorer (ACE) \citep{ACE}.
The database was also exploited to derive the geomagnetic indices $Kp$, $Dst$ and $Sym$-$H$ (at 3-hour, 1-hour, and 5-min resolutions, respectively), measured using ground-based ma\-gne\-to\-me\-ters.

Some minor geomagnetic storms occurred on 12 December, while the time interval preceding the shock was characterized by a relatively low geomagnetic activity.
The large increase in the SW velocity $V_{SW}$ on 14 December at $\sim$1410 UT, associated with the leading edge of the CME, caused the Storm Sudden Commencement (SSC).
The increased dynamical pressure $P_{SW}$ resulted in a dramatic magnetospheric compression along with an intensification of the magnetopause current.
The SSC, clearly visible in the time profiles of the geomagnetic indices, marked the beginning of the initial phase of the storm,
which was characterized by intense fluctuations in
$P_{SW}$
and in all IMF components.
In particular, $B_{z}^{IMF}$ became positive after 1800 UT on 14 December and it continued to oscillate until the $\sim$2300 UT, when the IMF intensity $B_{tot}^{IMF}$ increased and $B_{z}^{IMF}$ rapidly turned negative, while $V_{SW}$ decreased.
%Then the main phase of the storm began, reaching a maximum in the first hours of 15 December,
The main phase of the storm reached a maximum in the first hours of 15 December,
followed by a slow ($\sim$ 3 days) recovery phase.
The protracted large-amplitude (up to $\sim$18 nT) southward IMF was associated with the magnetic cloud which caused the storm \citep{KATAOKA}.
Such large events are untypical of the intervals of low solar activity.
An additional interplanetary shock, related to a less geo-effective CME, was registered on 16 December at $\sim$ 1800 UT.
Significant gaps in the ACE data are present after 1900 UT on 16 December.

\section{PAMELA's Observations}
PAMELA is a space-based experiment designed for a precise measurement of the charged
CRs (protons, electrons, their antiparticles and light nuclei)
in the kinetic energy range from some tens of MeV up to several hundreds of GeV \citep{PHYSICSREPORTS}.
The Resurs-DK1 satellite, which hosts the apparatus, was launched into a semi-polar (70 deg inclination) and elliptical (350--610 km altitude) orbit on 15 June 2006.
The instrument consists of a magnetic spectrometer equipped with a silicon tracking system, a time-of-flight system shielded by an anticoincidence system, an electromagnetic calorimeter and a neutron detector \citep{PICOZZA}.
Details about apparatus performance, proton selection, detector efficiencies and experimental uncertainties can be found elsewhere (see e.g. \citet{SOLARMOD}).

PAMELA is providing comprehensive observations of the interplanetary \citep{SOLARMOD,ELECTRONS} and magnetospheric \citep{TRAPPED,ALBEDO} radiation in the near-Earth environment.
In particular, PAMELA is able to accurately measure the SEP events during solar cycles 23 and 24 \citep{SEP2006,MAY17PAPER}, including energetic spectra and pitch angle distributions \citep{BRUNO_ICRC_SEP} in a wide interval, bridging the low energy data by in-situ spacecrafts and the Ground Level Enhancement (GLE)
data by the worldwide network of neutron monitors.

\subsection{Magnetic Coordinates}\label{Magnetic Coordinates}
Data were analyzed in terms of Altitude Adjusted Corrected GeoMagnetic (AACGM) coordinates, developed to provide a realistic description of high latitude regions by accounting for the multipolar geomagnetic field. They are defined such that all points along a magnetic field line have the same geomagnetic latitude and longitude, so that they are closely related to invariant magnetic coordinates \citep{BAKER, GUSTAFSON, HERES}. The AA\-CGM system coincides with the standard Corrected GeoMagnetic (CGM) system \citep{BREKKE} at the Earth's surface.
The computation is based on the IGRF-11 model \citep{IGRF11}, which employs a global spherical harmonic implementation of the main magnetic field;
AACGM latitudes at PAMELA's orbit are not significantly affected by the inclusion of external geomagnetic sources.

Alternatively, following the standard approach, cutoff observations were expressed as a function of the McIlwain's $L$ parameter \citep{MCILWAIN}.
In a dipole, $L$ numerically approximates the radius (measured in Re) where a geomagnetic field line crosses the equator.
The invariant latitude $\Lambda_{inv}$ of a location,
derived from $L$ by the relation: $cos\Lambda_{inv}$ = $L^{-1/2}$, represents a particularly useful parameter for the investigation of high latitude phenomena.

\subsection{Evaluation of Geomagnetic Cutoff Latitudes}
The lowest magnetic latitude to which a charged CR particle can penetrate the Earth's ma\-gne\-tic field is known as its \emph{cutoff latitude} and is a function of the particle
rigidity ($R$ = momentum/charge).
Alternatively, a \emph{cutoff rigidity} can be associated with a given location in the magnetosphere, corresponding to the minimum rigidity needed to access to the considered position.
Some complications arise from the presence of the Earth's solid body (together with its atmosphere):
both ``allowed'' and ``forbidden'' bands of CR particle access are present in the so-called ``penumbra'' region \citep{Cooke}.
Cutoff values are in general a function of
particle direction of arrival and geomagnetic activity.
Due to the narrow ($\sim$20 deg) field of view of PAMELA, with its major axis mostly oriented toward the zenith,
the measured CR fluxes correspond to approximately vertical directions.

The algorithm used to evaluate cutoff latitudes from the PAMELA data \citep{BRUNO_ICRC_CUTOFF} is si\-mi\-lar to one developed by \citet{LESKE2001},
using the low-energy proton and alpha particle measurements made by the Solar Anomalous and Magnetospheric Particle Explorer (SAM\-PEX) spacecraft.
For each rigidity bin, a mean flux was obtained by averaging fluxes measured at latitudes higher than $\Lambda_{min}=cos^{-1} (R[GV]/20)^{1/4}$ deg, and the cutoff latitude was evaluated as the latitude where the flux
intensity is equal to the half of the average value.
$\Lambda_{min}$, employed to improve the statistics at high rigidities, represents a rigidity dependent upper cutoff developed
to avoid penumbral effects,
and it was derived by using proton data acquired by PAMELA during relatively quiet geomagnetic conditions \citep{ALBEDO}.

To support the analysis results, cutoff latitudes were also numerically modeled with back-tracing techniques \citep{BRUNO_ICRC_SEP,BRUNO_ICRC_UNDERCUTOFF}.
Using the spacecraft ephemeris data, and the particle rigidity and direction provided by the tracking system,
trajectories of all selected protons were reconstructed in the Earth's magnetosphere by means of a tracing program based on \citet{TJPROG} and implementing a realistic semi-empirical description of internal and external geomagnetic field sources (see Section \ref{Comparison with Modeled Cutoffs}).
In order to discard geo\-ma\-gne\-tically trapped protons \citep{TRAPPED} and low energy albedo protons \citep{ALBEDO} in the equatorial regions,
while saving significant computational time, only protons with rigidities higher than 10/$L^{3}$ GV were selected. Trajectories were back propagated from the measurement location until they escaped the model magnetosphere boundaries (interplanetary CRs) or they reached an altitude of 40 km (albedo CRs).
Then, at a given rigidity, the \emph{modeled} cutoff latitude was evaluated as the latitude where the interplanetary and albedo flux intensities were equal.

Accounting for
the statistical limitations,
the calculation was performed for 18 rigidity logarithmic bins covering the interval 0.39--7.47 GV
and the final cutoff values were derived by
fitting PAMELA's observations averaged over single orbital periods ($\sim$94 min), including two cutoff measurements (entering and exiting the polar caps) in both magnetic hemispheres.
The relationships used to fit cutoff latitude data were obtained by inverting the formulas developed to describe cutoff rigidities:
\begin{equation}\label{Rcutoff_Lambda}
\centering
R(\Lambda) = a_{\Lambda}\cdot cos^{4}\Lambda - b_{\Lambda},
\end{equation}
\begin{equation}\label{Rcutoff_L}
R(L) = a_{L} / L^{2} - b_{L},
\end{equation}
as a function of AACGM (or invariant) latitude $\Lambda$ and $L$-shell respectively, with ($a_{\Lambda}, b_{\Lambda}$) and ($a_{L}, b_{L}$)
the corresponding fitting parameters \citep{ALBEDO}.
Such a pa\-ra\-me\-te\-ri\-za\-tion, which takes into account the magnetospheric effects at high latitudes, was introduced by \citet{OGLIORE} to reproduce CR nuclei observations made by the SAMPEX mission.
In general, the fit results can vary up to a factor 10\% at equatorial and mid-latitude locations due to the non-dipole terms in the geomagnetic field \citep{SMART2005}.

\subsection{Results}
\begin{figure*}[!t]
\centering
\begin{tabular}{c}
\includegraphics[width=5.6in]{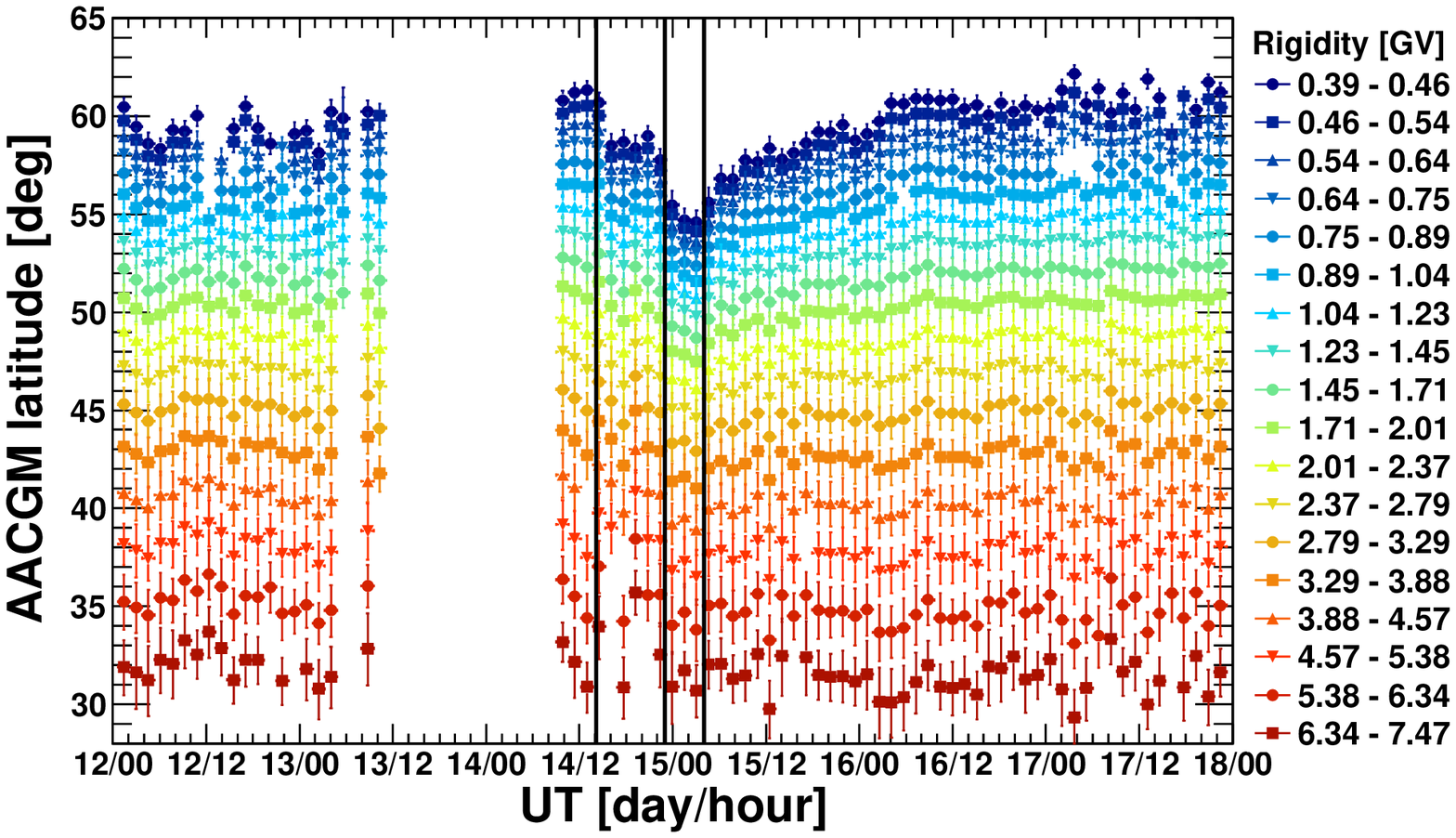} \\
\includegraphics[width=5.6in]{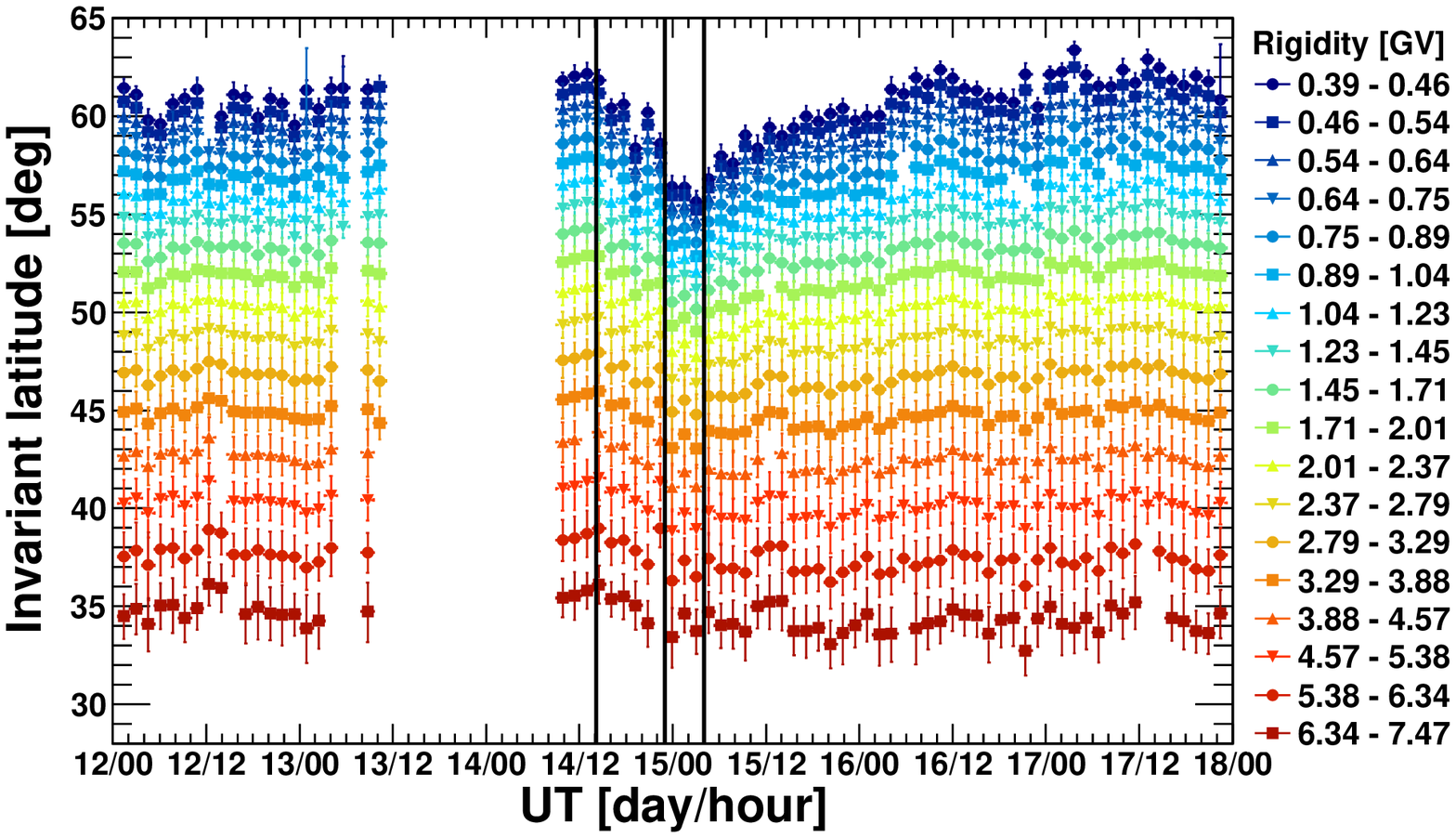}
\end{tabular}
\caption{Time profile of the geomagnetic cutoff latitudes measured by PAMELA, for different rigidity bins (color code). Cutoff results are reported for both AACGM (top) and invariant (bottom) latitudes. The vertical lines mark the beginning of the storm initial, main and recovery phases, respectively.}
\label{cTimeVarFit_PAM_AACGLat_iR20_iAACGL20_k2}
\end{figure*}

Figure \ref{cTimeVarFit_PAM_AACGLat_iR20_iAACGL20_k2} shows the geomagnetic cutoff latitudes measured by PAMELA for different rigidity bins (color code), between 12 and 18 December 2006.
Cutoff results are reported for both AACGM (top) and invariant (bottom) latitudes.
Each point denotes the cutoff latitude value averaged over a single spacecraft orbit (2 full polar passes); the error bars include the statistical uncertainties of the measurement.
In a few of cases
data points are missing because the numerical algorithm used to evaluate cutoff latitudes returned no cutoff or a bad cutoff value,
due to limited statistics.
The gap in the PAMELA data from 1000 UT on 13 December until 0914 UT on 14 December was related to an onboard system reset of the satellite.

The evolution of the 14 December magnetic storm followed the typical scenario in which the
cutoff latitudes move equatorward as a consequence of a CME impact on the magnetosphere with an associated transition to southward $B_{z}^{IMF}$.
The maximum cutoff suppression was observed during the storm's main phase at about 0300 UT on 15 December.
The registered deviation with respect to quiet geomagnetic conditions decreases with increasing rigidity,
with a $\sim$7 deg maximum suppression for the lowest rigidity bin (0.39--0.46 GV).
%Data supporting Figure \ref{cTimeVarFit_PAM_AACGLat_iR20_iAACGL20_k2} are available in the Supporting Information Table S1.

\begin{figure*}[!t]
\centering
\includegraphics[width=6.4in]{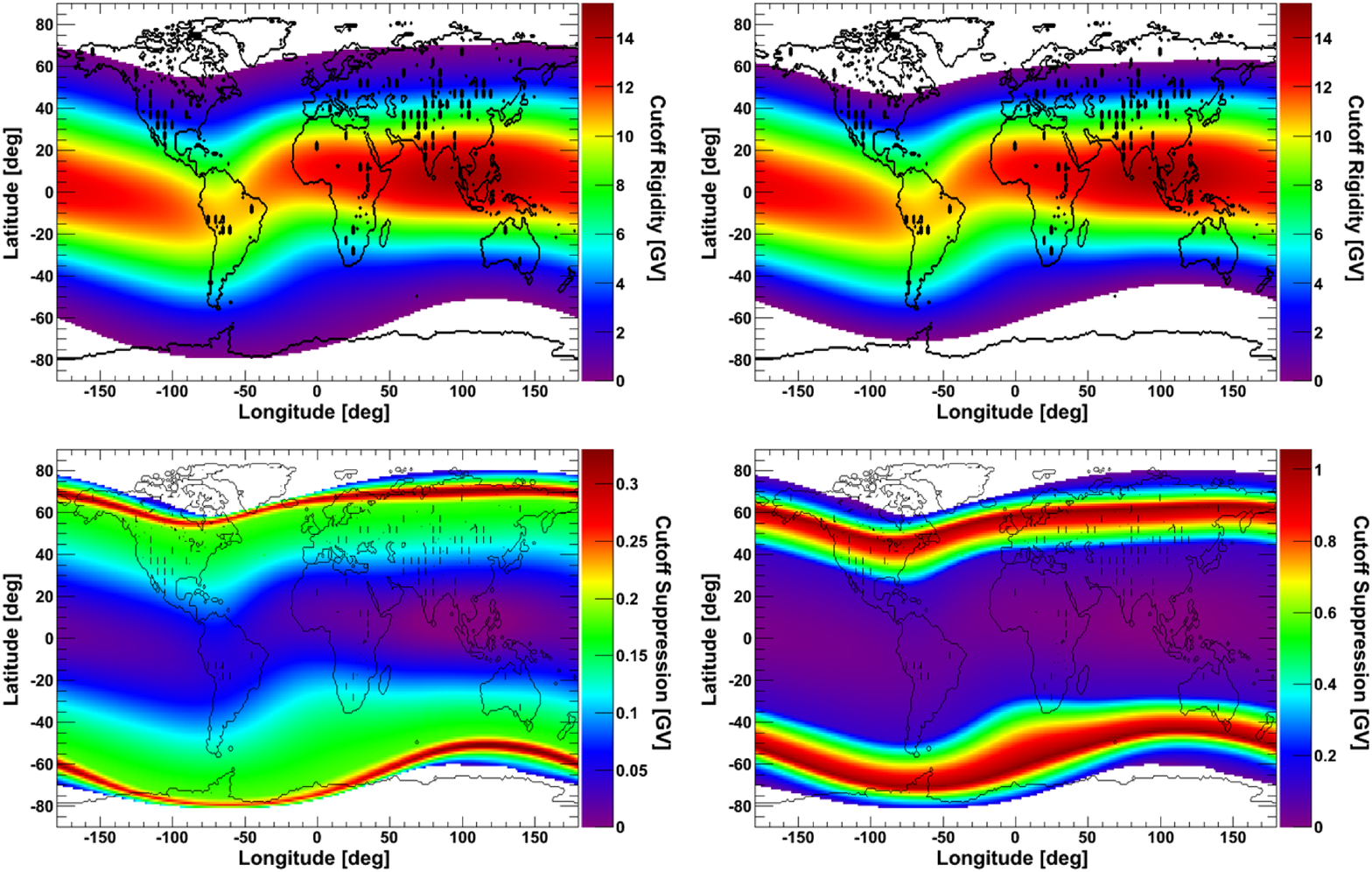}
\vspace{0.1cm}
\caption{Cutoff rigidity maps evaluated at the shock arrival (top-left), and at the time of maximum cutoff suppression (top-right).
Bottom panels show the corresponding cutoff decrease with respect to geomagnetically quiet conditions.}
\label{cutoff_maps}
\end{figure*}

Globally mapped cutoffs in geographic coordinates at mean PAMELA altitude (475 km) are shown in Figure \ref{cutoff_maps}.
The maps were derived from fitted results by evaluating cutoff rigidities (extrapolated down to 0 GV) as a function of $L$ through Equation \ref{Rcutoff_L}, and averaging PAMELA data over the two hemispheres and over the longitudes.
The top-left panel displays cutoffs measured on 14 December during the orbital interval 1344--1518 UT, including the shock arrival on the magnetosphere; the top-right panel report the results on 15 December between 0216 and 0350 UT, at the maximum registered cutoff suppression.
Bottom panels show the corresponding cutoff differences with respect to geomagnetically quiet conditions.
Significant discrepancies were found in both cases:
the rigidity cutoff suppression after the shock arrival is $\sim$0.33 GV at highest magnetic latitudes, about a third of the maximum deviation registered in the main phase of the storm ($\sim$1 GV).

\begin{figure*}[!t]
\centering
\includegraphics[width=6.4in]{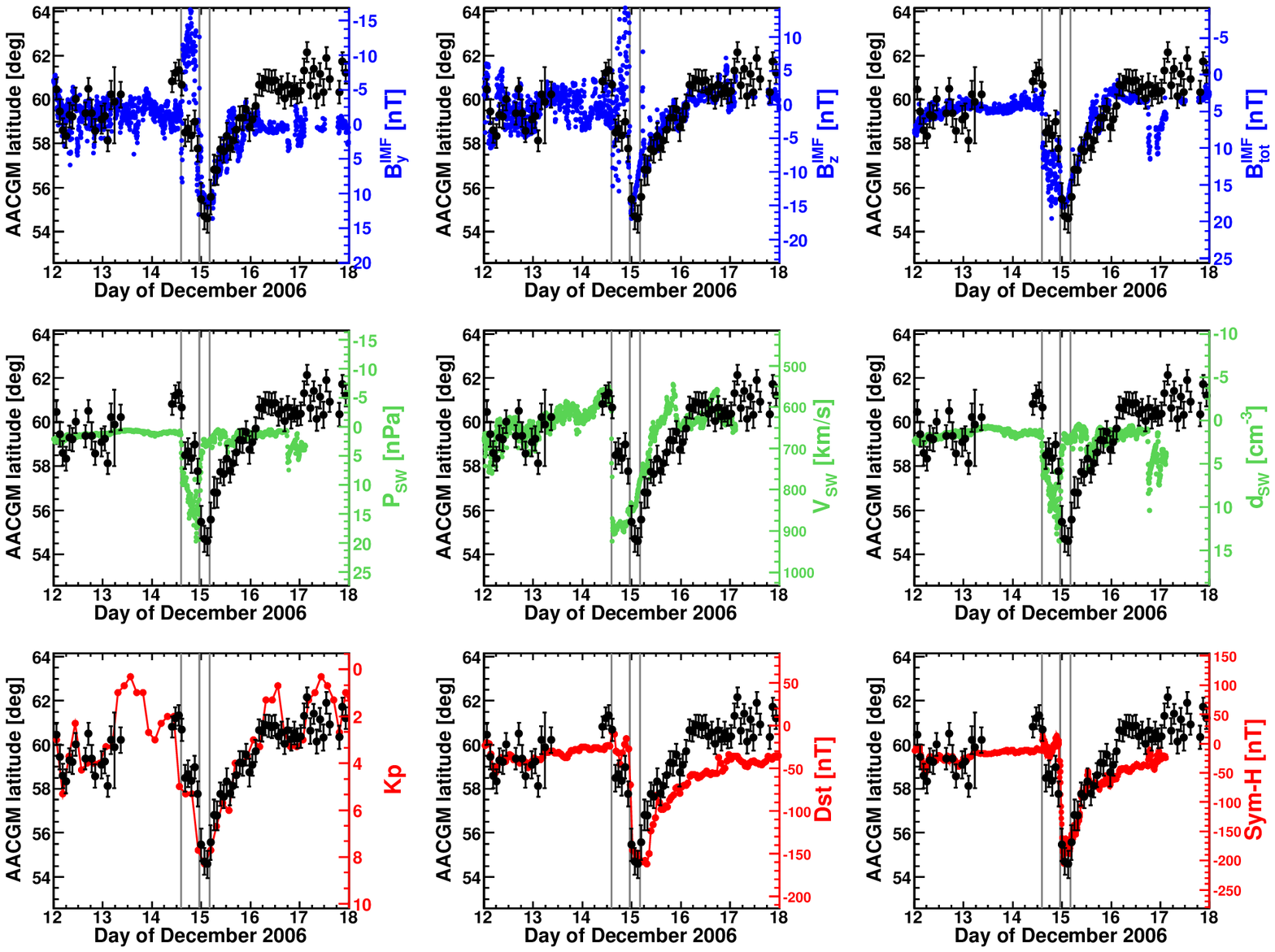}
%\vspace{0.1cm}
\caption{Time profiles of the IMF ($B_{y}^{IMF}$, $B_{z}^{IMF}$ and $B_{tot}^{IMF}$; blue), solar wind ($P_{SW}$, $V_{SW}$ and $d_{SW}$; green) and geomagnetic ($Kp$, $Dst$, $Sym$-$H$; red) parameters, compared with the variations in measured cutoff latitudes (0.39--0.46 GV; black). The vertical lines mark the beginning of the storm initial, main and recovery phases, respectively.}
\label{time_comp}
\end{figure*}

\subsubsection{Correlations with Interplanetary and Geomagnetic Parameters}
\begin{figure*}[!t]
\centering
\includegraphics[width=6.4in]{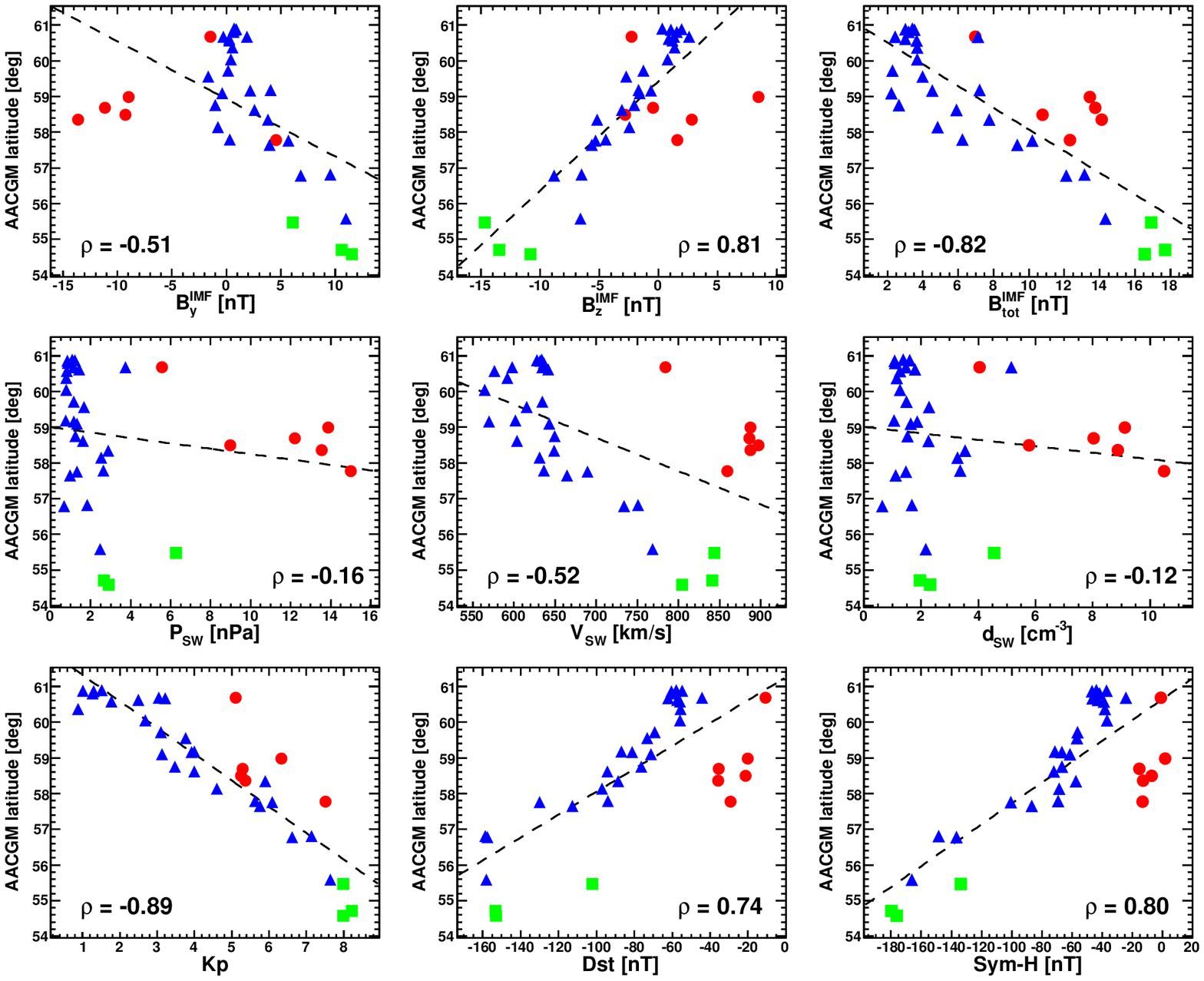}
%\vspace{0.1cm}
\caption{Correlation between measured cutoff latitudes in the 0.39--0.46 GV rigidity bin (Y-axis) and main IMF, SW and geomagnetic parameters (X-axis).
Orbital-averaged data during the three (initial, main, recovery) storm phases are indicated by red circles, green squares and blue triangles, respectively.
The corresponding total correlation coefficient $\rho$ is reported in each panel.}
\label{corr_plots}
\end{figure*}

Figure \ref{time_comp} reports the variations in the measured cutoff latitudes in the lowest rigidity bin (0.39--0.46 GV), compared to the time profiles of IMF ($B_{y}^{IMF}$, $B_{z}^{IMF}$ components and $B_{tot}^{IMF}$ total intensity), SW (dynamical pressure $P_{SW}$, velocity $V_{SW}$ and density $d_{SW}$) and geomagnetic ($Kp$, $Dst$ and $Sym$-$H$ indices) parameters. Correlation coefficients were estimated by interpolating and averaging Omniweb data (characterized by different time resolutions) over PAMELA orbital periods ($\sim$94 min).
The results of the analysis of the correlations are shown in Figure \ref{corr_plots}.
Partial correlation values for the three (initial, main, recovery) storm phases are reported in Table \ref{table_corr}, along with the total correlation coefficients.
Values in bold correspond to a strong correlation ($\geq$0.80) or anti-correlation ($\leq$-0.80).

During the initial phase (14 December, 1410--2300 UT), the most geo-effective pa\-ra\-me\-ters were $P_{SW}$, $d_{SW}$ and $B_{tot}^{IMF}$; a good correlation was observed for $Dst$ and $Sym$-$H$.
In the main phase (lasting up to $\sim$ 0400 UT on 15 December), very pronounced inter\-pla\-ne\-ta\-ry effects were exerted by $B_{y}^{IMF}$, $B_{tot}^{IMF}$, $P_{SW}$ and $d_{SW}$, while $B_{z}^{IMF}$ was moderately correlated.
Note that SW parameters were anti-correlated and correlated with the cutoff latitude variations in the initial and the main phases, respectively.
Furthermore, a significant correlation was shown by $Dst$ and $Sym$-$H$; however, the $Dst$ index was characterized by a very slow decrease between 0300--0700 UT on 15 December, and the maximum cutoff suppression in PAMELA data was observed about 4 hours before the $Dst$ minimum (-162 nT).
Instead, $Kp$ changes were much less correlated due to the coarser resolution.
Finally, the recovery phase was characterized by a strong correlation with $B_{tot}^{IMF}$ and, especially, with $B_{z}^{IMF}$ and all the three geomagnetic indices.
$B_{y}^{IMF}$ and $V_{SW}$ were also quite geo-effective.
Note that the time period considered for the recovery phase is limited to 1900 UT on 16 December, due to the gaps in ACE data.
In general,
as demonstrated in Figure \ref{corr_plots},
the shapes of the time variations in the PAMELA's cutoff measurements were strongly correlated with $B_{z}^{IMF}$, $B_{tot}^{IMF}$, $Sym$-$H$ and especially $Kp$.
A less significant correlation was observed for $Dst$:
while the $Kp$ changes appeared to lead the cutoff suppressions,
$Dst$ was found to respond with some delay (see Figure \ref{time_comp}), in agreement with previous studies \citep{LESKE2001,NEAL2013}.

\begin{table}[!t]
\centering
\begin{tabular}{|c||c|c|c|c|c|c|c|c|c|}
  \hline
  Phase  & $B^{IMF}_{y}$ & $B^{IMF}_{z}$ & $B^{IMF}_{tot}$ & $P_{SW}$ & $V_{SW}$ & $d_{SW}$ & $Kp$ & $Dst$ & $Sym$-$H$ \\
  \hline
  \hline
Initial & 0.03 & -0.28 & -0.70 & -0.76 & -0.41 & -0.74 & -0.56 & 0.75 & 0.73 \\
Main & \textbf{-0.95} & 0.68 & \textbf{-0.93} & \textbf{0.93} & 0.35 & \textbf{0.94} & -0.69 & \textbf{0.89} & \textbf{0.93} \\
Recovery & -0.74 & \textbf{0.94} & \textbf{-0.84} & -0.28 & -0.73 & -0.08 & \textbf{-0.94} & \textbf{0.94} & \textbf{0.91} \\
  \hline
All & -0.51 & \textbf{0.81} & \textbf{-0.82} & -0.16 & -0.52 & -0.12 & \textbf{-0.89} & 0.74 & \textbf{0.80} \\
  \hline
\end{tabular}
\vspace{0.4cm}
\caption{Correlation coefficients between measured cutoff latitudes (0.39--0.46 GV) and main IMF, SW and geomagnetic parameters, during the three different (initial, main, recovery) storm phases and the whole storm (all).}
\label{table_corr}
\end{table}

\subsubsection{Comparison with Modeled Cutoffs}\label{Comparison with Modeled Cutoffs}
\begin{figure*}[!t]
\centering
\includegraphics[width=4.8in]{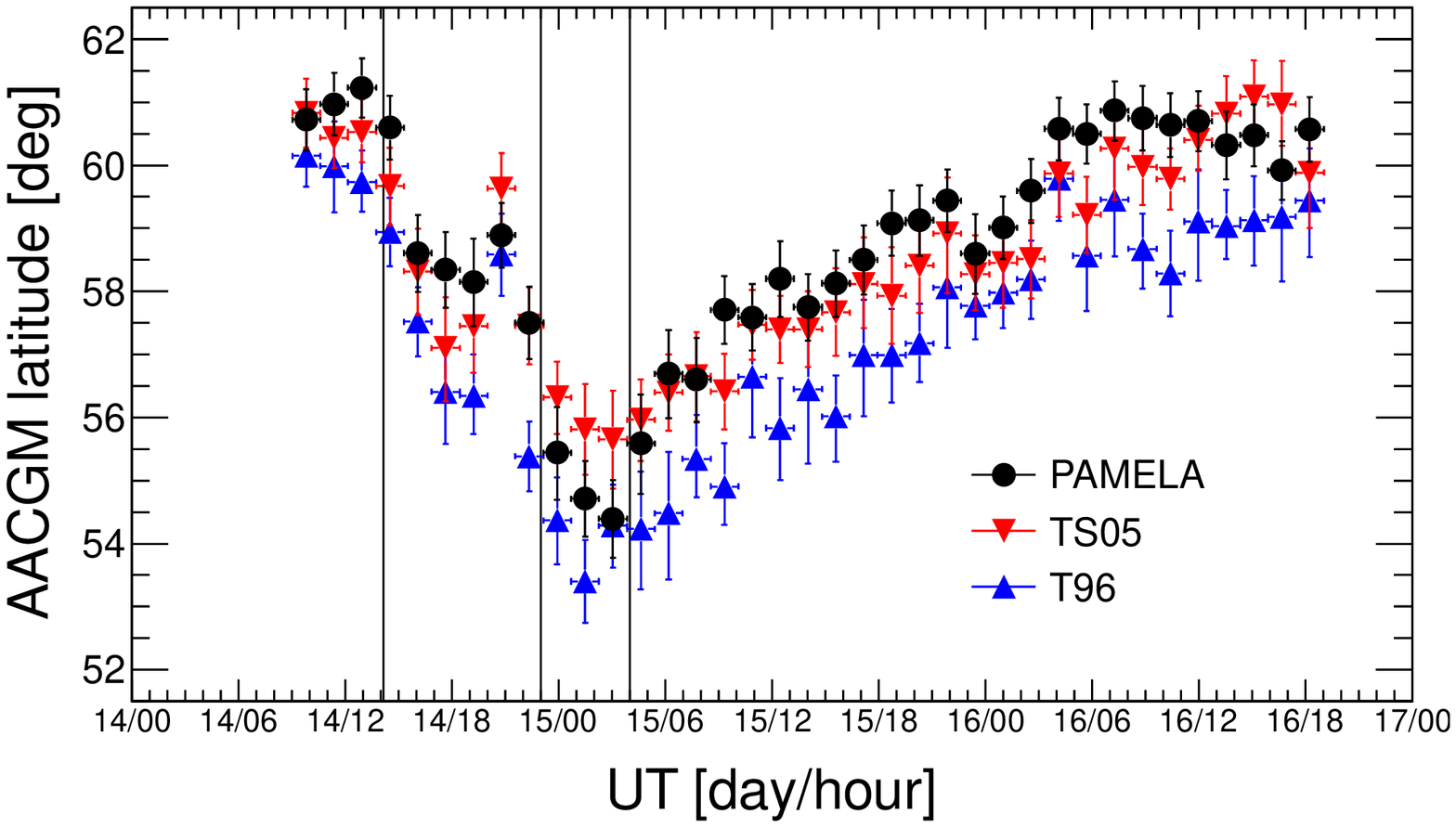}
%\vspace{0.1cm}
\caption{Comparison between measured (black circles) and modeled (blue triangles - T96 model; red triangles - TS05 model) cutoff variations in the lowest rigidity interval: 0.39--0.46 GV. The vertical lines mark the beginning of the storm initial, main and recovery phases, respectively.}
\label{comparison}
\end{figure*}

Figure \ref{comparison} reports the comparison between measured and modeled cutoff latitudes during the geomagnetic storm, for the lowest rigidity bin (0.39--0.46 GV).
Modeled cutoffs were obtained by using the IGRF-11 and the \citet{TS05} (TS05) models for the description of internal and external geomagnetic field sources, respectively. The TS05 model is a high resolution dynamical model of the storm-time magnetosphere,
based on recent satellite measurements.
The model input consists in
$B_{y}^{IMF}$, $B_{z}^{IMF}$, $P_{SW}$ and $Dst$; 6 additional supplied parameters describe the strength of the symmetric ring current, the partial ring current, the Birkeland 1 and 2 currents, and two different tail current systems. Alternatively, the less sophisticated \citet{T96} (T96) model was used.

PAMELA data (black) are well reproduced by the TS05 model (red) within statistical errors:
on average, modeled cutoff latitudes are $\sim$0.31$\pm$1.22 deg equatorward shifted;
small mean differences
can be observed in the initial and recovery phases of the storm (about -0.32$\pm$1.25 deg and -0.43$\pm$1.20 deg, respectively), while the modeled cutoffs are $\sim$1.07$\pm$1.34 deg poleward in the main phase; overall, the maximum deviation from PAMELA's cutoffs is $\sim$1.3 deg.

On the other hand, the cutoff latitudes estimated with the T96 model (blue) are syste\-ma\-ti\-cally ($\sim$1.49$\pm$1.30 deg) equatorward of the PAMELA's observations during all storm phases, with a larger maximum deviation ($\sim $2.8 deg). This is not unexpected since the magnetospheric description provided by the T96 model is not adequate in the case of intense geomagnetic activity, overestimating storm effects \citep{DESORGHER}.

\section{Summary and Conclusions}

In this study we have taken advantage of the proton data of the PAMELA satellite experiment to perform a measurement of the geomagnetic cutoff variations during the long lasting
storm on 14 December 2006. The arrival of the 13 December CME caused a strong perturbation of the local radiation environment, affecting the planetary CR distribution.
The evolution of the consequent  geomagnetic storm followed the typical scenario in which the cutoff latitudes move equatorward as a consequence of a magnetic cloud impact on the Earth's magnetosphere with an associated transition to southward IMF.
A significant reduction in the geomagnetic shielding was observed, with a maximum cutoff latitude suppression of about 7 deg at lowest rigidities.
Such large CME-driven storms are relatively rare during the intervals of low solar activity.
The variability of the cutoff latitude as a function of rigidity was studied on relatively short timescales, corresponding to single spacecraft orbits ($\sim$94 min).
Measured cutoff variations were related to the changes in the magnetosphere configuration, investigating the role of IMF, SW and geomagnetic parameters.
In particular, we found a high correlation with the variations of $B_{z}^{IMF}$, $B_{tot}^{IMF}$ and the geomagnetic activity as measured by the $Kp$ index and, to a lesser extent, by the $Sym$-$H$ index.
Finally, results were compared with those obtained with back-tracing techniques based on a realistic semi-empirical modeling of the magnetosphere.
PAMELA's observations represent the first direct measurement of geomagnetic cutoffs for protons with kinetic energies from $\sim$ 80 MeV to several GeV.

\begin{acknowledgments}
We acknowledge support from The Italian Space Agen\-cy (ASI), Deutsches Zentrum f\"{u}r Luftund Raumfahrt (DLR), The Swedi\-sh National Space Board, The Swedi\-sh Research Council, The Russian Space Agency (Ros\-cosmos) and The Russian Scientific Foundation.
\end{acknowledgments}

%\end{article}
\end{document}